\documentclass{ws-ijgmmp}

\usepackage{amssymb}
\usepackage{amsmath}

\usepackage{empheq}
\usepackage{comment}
\usepackage{mathrsfs}
\usepackage{calligra}

\DeclareMathOperator{\Wnabla}{\overset{\rm w}{\nabla}}
\DeclareMathOperator{\Christo}{\Gamma}
\DeclareMathOperator{\WChrist}{\overset{\rm w}{\Gamma}}

\begin{document}
\markboth{T. Wada, S. Noda}
{WEYL SYMMETRY OF THE GRADIENT-FLOW IN INFORMATION GEOMETRY }

%
\catchline{}{}{}{}{}
%
\title{WEYL SYMMETRY OF THE GRADIENT-FLOW IN INFORMATION GEOMETRY 
}


\author{TATSUAKI WADA}

\address{ Region of Electrical and Electronic Systems Engineering, Ibaraki University, \\
4-12-1 Nakanarusawa-cho, Hitachi, Ibaraki, 316-8511, Japan\\
\email{tatsuaki.wada.to@vc.ibaraki.ac.jp
} }

\author{SOUSUKE NODA}
\address{Miyakonojo College, National Institute of Technology, \\
 473-1, Yoshiocho, Miyakonojo, Miyazaki, 885-8567, Japan\\
 \email{snoda@cc.miyakonojo-nct.ac.jp}
 }
 
 \maketitle
 
 \begin{history}
\received{(Day Month Year)}
\revised{(Day Month Year)}
\end{history}

\begin{abstract}
We have revisited the gradient-flow in information geometry from the perspective of Weyl symmetry.
The gradient-flow equations are derived from the proposed action which is invariant under the Weyl's gauge transformations.
In Weyl integrable geometry, we have related Amari's $\alpha$-connections in information geometry to
the Weyl invariant connection on the Riemannian manifold equipped with the scaled metric.
\end{abstract}


\section{Introduction}
\label{intro}
Weyl geometry \cite{Weyl, PS14} is one of the simple generalizations of Riemannian geometry, 
and it provides a natural framework in which scale invariance is realized
as a local symmetry. In Weyl geometry, a parallel transported vector changes its length, which is characterized by the \textit{non-metricity}
\begin{align}
  \Wnabla_{\rho} \, g_{\mu \nu} = \omega_{\rho} \, g_{\mu \nu}, \quad  \Wnabla_{\rho} \, g^{\mu \nu} = -\omega_{\rho} \, g^{\mu \nu},
  \label{non-metricity}
  \end{align}
where $\omega_{\rho}$ is a Weyl gauge field, or covariant vector (covector) field and $\Wnabla$ is a Weyl connection (see \eqref{Wconnection}).
H. Weyl \cite{Weyl} introduced the concept of so-called \textit{Weyl symmetry}.
A theory or equation, which consists of a metric $g_{\mu \nu}(x)$ and a scalar field $\varphi(x)$, possess Weyl symmetry, if it is invariant under the Weyl transformation,
\begin{align}
  g_{\mu \nu}(x) \to \Omega^2(x) \, g_{\mu \nu}(x), \quad \varphi(x) \to \Omega^{-1}(x) \, \varphi(x), \quad \omega_{\mu} \to \omega_{\mu} + \partial_{\mu} \ln \Omega^2(x),
  \label{Weyl-transform}
\end{align}
which make a local rescaling of $g_{\mu \nu}(x)$, $\varphi(x)$, and Weyl  gauge field $\omega_{\mu}$ with a scalar function $\Omega(x)$.
The Weyl symmetry is defined by the equivalent classes of $( g_{\mu \nu}, \omega_{\rho} )$. Each equivalent class is related to another by the Weyl gauge transformation \eqref{Weyl-transform}.
Choosing a certain equivalent class of $( g_{\mu \nu}, \omega_{\rho} )$ is called the \textit{gauge fixing}.
Different choices of  $( g_{\mu \nu}, \omega_{\rho} )$ lead to different Weyl gauges. A special class of $( g_{\mu \nu}, \omega_{\rho} = 0 )$ is called Einstein gauge, in which the non-metricity \eqref{non-metricity} reduce to the metricity in Riemannian geometry and a Weyl gravitational theory \cite{RFP} reduces to the standard theory of general relativity (GR) proposed by Einstein.

In Riemannian geometry, it is well known that
the motion of a spinless particle is described by the action
$S_a = - m \int d s$,
where the constant $m$ denotes the mass of a particle and $ds$ is the line-element such that
$ds^2 = g_{\mu \nu} dx^{\mu} dx^{\nu}$.
In contrast, in Weyl geometry, the mass parameter, which is the length of the 4-momentum vector $(p^2 = -m^2$),
is no longer a constant but it is a function of the spacetime point, i.e., $m= m(x)$.
The action principle is modified 
as
\begin{align}
  S_a = - \int m(x) d s,
  \label{invariant-action}
\end{align}
which is invariant under the scale transformation \eqref{Weyl-transform} of $ds \to \Omega(x) \, ds$ and $m(x) \to m(x) / \Omega(x)$ with a positive scalar function $\Omega(x)$.

On another note,
information geometry (IG) \cite{AN,Amari} is a useful method for exploring the fields of information science by means of differential geometry.
IG is invented from the studies on invariant properties of a manifold of probability distributions.
An elementary introduction to IG is provided by Nielsen \cite{Nielsen}, and Caticha's short paper \cite{Caticha} enlightens the basics of IG by
introducing the "distance" between two probability distributions.
It is known that the gradient-flow equations are useful for some optimization problems, e.g.,  the gradient-flow in  metric spaces \cite{AGS08}.
The gradient flows on a Riemann manifold follow the direction of gradient descent (or ascent) in the landscape of a potential functional, with respect to (w.r.t.) the curved structure of the underlying metric space. 
Since Fujiwara and Amari \cite{FA95} studied the information geometric studies on the gradient systems,
the several studies on the gradient-flow in IG have been done from the different perspectives.
 Ref.~\cite{MP15} studied the natural gradient-flow in the mixture geometry of a discrete exponential family.
Ref.~\cite{BN16} studied the relationship between Hamiltonian flow and gradient flow from the perspective of symplectic geometries. 
One of the authors and his collaborators have been studied the gradient-flow in IG extensively, and related them with some different fields in physics.
For examples, Ref. \cite{WSM21,WSM22} studied this issue from the perspective of geometric optics and showed that a path of the gradient-flow can be regarded as the light path (or ray) described by the anisotropic Huygens equation.
The gradient-flow are also related to the thermodynamic processes, and it is shown \cite{WSM21} that the evolutional parameter in the gradient flow equations in IG is related to the temperature of the simple thermodynamic systems based on the Hamilton-Jacobi dynamics.
The gradient-flow are related to the co-geodesic flow of the geodesic Hamiltonians \cite{WSM22}.
Furthermore the analytical mechanical properties and thermodynamics of black holes concerning the gradient-flow are studied in \cite{CW22}, and 
discussed  the deformations of the gradient-flow equations by using Randers-Finsler metrics. 
Through a series of these studies, it is important that treating space and time on equal footing, which is an essence of Einstein's relativity. In order to conveniently describe physical equations in GR, the selection of coordinates system in curved spacetime is important and non-trivial. We believe that the same as true for describing the gradient-flow in IG.
Ref. \cite{W23} relates the gradient-flow in IG and Weyl integrable geometry.
Ref. \cite{WS24} studied the gradient-flow in IG from the motion of a null (or light-like) particle in a curved space, and derived the Hamiltonians which describe the gradient-flow.

In this work, we reconsider the gradient-flow equations in IG from the perspective of a time-like particle in a curved spacetime, in which
Weyl symmetry plays a fundamental role.
It is shown that the pre-geodesic equations (non-affinely parametrized geodesic equations) associated with the gradient-flow in IG are related to the autoparallel equations in the Weyl integrable geometry. Weyl's gauge symmetry plays a significant role. As a result, we found that Weyl's gauge transformations relate the $\alpha$-connection with $\alpha \!=\! \pm 1$ to the Levi-Civita connection on the Riemannian manifold equipped with the scaled metric, which consists of Fisher metric and a scalar function.
The rest of paper is organized as follows. Section 2 reviews the Weyl integrable geometry. Wely's non-metricity conditions, the autoparallel equations, and Weyl's gauge transformations are explained.
Section 3 provides some basics of IG and the gradient-flow equations in IG.
Section 4 is the main section. After some brief explanations on the motion of a particle with position dependent mass $m(x)$ in a curved spacetime, we derive the gradient-flow equations in IG from the Euler-Lagrange (EL) equations of the action which has the Weyl symmetry.
Final section is devoted to our conclusions.
In \ref{app1} we provide the derivation of the Weyl autoparallel equations, and in \ref{app2} we explain the equivalence between
the autoparallel and geodesic equations.

Throughout the paper, we use Einstein's summation convention for repeated indices.
Latin indices run over space, e.g. $i, j = 1,\ldots, N$, while Greek indices run over spacetime, e.g. $\mu, \nu = 0,1, \ldots, N$.
The sign convention (the signature of the metric) is chosen as $(-, \underbrace{+,\ldots, +}_{N})$.

\section{Weyl Integrable Geometry}
Only a few years after Einstein proposed his theory of general relativity (GR), Weyl \cite{Weyl} introduced the concept of Weyl symmetry.
Weyl geometry $(\mathcal{M}, g, \omega_{\rho})$  \cite{PS14,RFP} is his generalization of Riemannian geometry. Weyl introduced a gauge field, or covector field $\omega_{\rho}$ 
and the \textit{non-metricity conditions} \eqref{non-metricity} are assumed.
When $\omega_{\rho} = 0$ the non-metricity \eqref{non-metricity} reduce to the metricity $\nabla g = 0$  in Riemannian geometry.
When a Weyl covector $\omega_{\rho}$ is a gradient of a scalar field, say
$ \omega_{\rho}(x) = \partial_{\rho} \phi(x)$, it is called \textit{Weyl integrable geometry}.
In this case, in addition to the metric, a scalar field $\phi(x)$ is the fundamental object which characterizes Weyl integrable geometry.
In this paper we restrict ourselves in the Weyl integrable geometry.

As a similar way for obtaining the coefficients of Levi-Civita connection,
by using the non-metricity conditions  \eqref{non-metricity} we can obtain 
the coefficients $\WChrist{}^{\mu}{}_{\nu \rho}(x)$ of Weyl's connection $\Wnabla$ \cite{PS14},
\begin{align}
 \WChrist{}^{\mu}{}_{\nu \rho}(x) = \Christo^{\mu}{}_{\nu \rho} (x) - \frac{1}{2} \left\{\delta^{\mu}_{\nu} \, \partial_{\rho} \phi(x) + \delta^{\mu}_{\rho}  \, \partial_{\nu} \phi(x) \, 
   - g^{\mu \sigma}(x)  \, \partial_{\sigma} \phi(x)  \,  g_{\nu \rho}(x)\right\},
   \label{Wconnection}
 \end{align}
where $\Christo^{\rho}{}_{\mu \nu}(x)$ denotes the coefficients of Levi-Civita connection w.r.t. a metric $g$.

For any smooth curve $C=C(\lambda)$ with a parameter $\lambda$, and any pair of two parallel vector fields $V$ and $U$ along $C$, 
we have
\begin{align}
  \frac{d} {d \lambda}  \left( g_{\mu \nu} V^{\mu} U^{\nu} \right) =  \partial_{\rho} \phi \, \frac{d x^{\rho}}{d \lambda} \, g_{\mu \nu} V^{\mu} U^{\nu},
\end{align}
 in a coordinate basis.
For a given Weyl covector $\omega_{\rho} = \partial_{\rho} \phi$ satisfying Weyl's non-metricity conditions \eqref{non-metricity}, the most general expression of the autoparallel equations (see \ref{app1}) are given by
\begin{align}
  \frac{d x^{\nu}}{d \lambda} \Wnabla_{\nu} \frac{d x^{\mu}}{d \lambda} = \frac{1}{2 u^2(\lambda)} \left( \frac{d u^2(\lambda)}{d \lambda} - u^2(\lambda) \, \partial_{\rho} \phi \, \frac{d x^{\rho}}{d \lambda}  \right) \frac{d x^{\mu}}{d \lambda},
  \label{WgeodesicEq}
\end{align}
where the tangent vector $u^{\mu}(\lambda) := d x^{\mu} / d \lambda$ and
\begin{align}
  u^2(\lambda) := g_{\mu \nu}(x) \frac{d x^{\mu}}{d \lambda} \frac{d x^{\nu}}{d \lambda}.
\end{align}
Note that the relation \eqref{non-metricity} is invariant under the Weyl gauge transformations \eqref{Weyl-transform}.
In addition, one can readily check that Weyl's connection \eqref{Wconnection} is invariant under the gauge transformations \cite{PS14, RFP}.

The proper time is generalized with the Weyl gauge field $\omega_{\rho}(x) = \partial_{\rho} \phi(x)$ as follows.
For a given tangent vector $v^{\nu}(\tau) := d x^{\nu} / d \tau$ and Weyl gauge (or covector) fileld $\omega_{\rho}(x) = \partial_{\rho} \phi(x)$, the parameter $\tau$ is called a \textit{proper time} \cite{ADR18} in Weyl geometry if it satisfies
 \begin{align}
    \frac{d v^2(\tau)}{d \tau} = v^2(\tau) \, \partial_{\rho} \phi (x) \, \frac{d x^{\rho}}{d \tau},
    \label{Proptime}
 \end{align}
 where $v^2(\tau) := g_{\mu \nu}(x) v^{\mu} (\tau) v^{\nu}(\tau)$.
 If a parameter $\lambda$ in \eqref{WgeodesicEq} is the proper time $\tau$, then the right-hand side (RHS) of  \eqref{WgeodesicEq} becomes zero,
 and consequently \eqref{WgeodesicEq} reduces to  the geodesic equation
 $( d x^{\nu} / d \tau ) \Wnabla_{\nu} (d x^{\mu} / d \tau)  =0$.

 \section{Information geometry and gradient-flow equations}
IG  \cite{AN,Amari,Caticha} is a differential geometry on a statistical manifold constructed from the parameterized probability distribution function (pdf) and the so-called \textit{dually-flat structure} is important.
IG is a non-Riemannian geometry equipped with Fisher metric $g$ and the dual $\alpha$-connections $\nabla^{(\alpha)}$, $\nabla^{\star (\alpha)}$.
These points are briefly explained here.

For a given set of some functions $F_i(x), i=1, \dots, N$, the $\theta$-parametrized pdf
\begin{align}
 p_{\theta}(x) = \exp \left[ \theta^i  F_i(x) - \Psi(\theta)  \right],
\end{align}
is called  \textit{exponential family}, where $x$ denotes an instanteneous value of the stochastic variable $\chi$.
Typical examples of exponential family are Gauss (or normal), gamma, beta, Bernoulli, Poisson pdfs.
Here $\Psi(\theta)$ is determined from the normalization of $p_{\theta}(x)$ as $\Psi(\theta) = \ln \left[ \int dx \exp (\theta^i F_i(x) ) \right]$.
$\Psi^\star(\eta) $ is the Legendre dual to $\Psi(\theta)$
\begin{align}
    \Psi^\star(\eta) = \theta^i \, \eta_i - \Psi(\theta), \quad \theta^i &= \frac{\partial \Psi^\star(\eta)}{\partial \eta_i}, \quad 
\eta_i = \frac{\partial \Psi(\theta)}{\partial \theta^i},
\end{align}
where $\theta^i$ and $\eta_i$ are called the dual affine coordinates. 
Since $ \eta_i = {\rm E}_{p_{\theta}} [ F_i(x) ]$, it is also called \textit{expectation coordinate}.
The $\eta$-potential function $\Psi^\star(\eta)$ is the (negative of ) entropy, i.e.,
\begin{align} 
 \Psi^\star(\eta) = {\rm E}_{p_{\theta}} [ \ln  p_{\theta}(x) ].
\end{align}

For these convex function $\Psi(\theta)$ and $\Psi^\star(\eta)$, one can construct the dually-flat structure as follows.
Firstly, the positive definite matrices $g_{ij}(\theta)$ and $g^{ij}(\eta)$ are obtained from the Hessian matrices
of the convex function $\Psi(\theta)$ and $\Psi^\star(\eta)$ as 
\begin{align} 
g_{ij} ( \theta) = 
\frac{\partial \eta_i}{\partial \theta^j} = 
\frac{\partial^2 \Psi( \theta)}{\partial \theta^i \partial \theta^j}, \quad
g^{ij} ( \eta) = 
\frac{\partial \theta^i}{\partial \eta_j} = 
\frac{\partial^2 \Psi^\star(\eta)}{\partial \eta_i \partial \eta_j}.
\label{g} 
\end{align}
These matrices satisfy the relation $g^{ij} (\eta) \,
g_{jk} (\theta) = 
\delta^i_k$,
where $\delta^i_k$ denotes Kronecker's delta.

Secondly, we introduce the $\alpha$-connections $\nabla^{(\alpha)}$ which is
a one-parameter ($\alpha$) extension $ \left\{ \nabla^{\alpha} \right\}_{\alpha \in \mathbb{R}}$ of Levi-Civita's connection and its dual $\nabla^{\star (\alpha)}$ as
\begin{subequations}
 \label{a-connection}
\begin{empheq}[left=\empheqlbrace]{align}
   \Christo^{(\alpha)}{}_{ijk}(\theta) &:=    \Christo_{ijk}(\theta) -\frac{ \alpha}{2} C_{ijk}(\theta) = \frac{(1- \alpha)}{2} \, C_{ijk}(\theta),  \\
   \Christo^{\star (\alpha) \; ijk} (\eta) &:= \Christo^{ijk}(\eta)  + \frac{\alpha}{2} {C^{\star}}^{ijk}(\eta) =  \frac{(1+ \alpha)}{2} \, {C^{\star}}^{ijk}(\eta),
\end{empheq}
\end{subequations}
where $ \Christo_{ijk}(\theta)$ ($ \Christo^{ijk}(\eta)$ ) are the Christoffel symbols of the first kind w.r.t. $g_{ij}(\theta)$ ($g^{ij}(\eta)$).
Here $C_{ijk}(\theta)$ and ${C^{\star}}^{ijk}(\eta)$ are the total symmetric cubic tensors, which are called \textit{Amari-Chentsov} tensors
\begin{align}
   C_{ijk}(\theta)  := \frac{\partial^3 \Psi(\theta)}{\partial \theta^i \partial \theta^j \partial \theta^k}, \quad
   {C^{\star}}^{ijk}(\eta)  := \frac{\partial^3 \Psi^\star(\eta)}{\partial \eta_i \partial \eta_j \partial \eta_k}.
\end{align}
Since the connection coefficients $ \Christo^{(\alpha)}_{ijk}$ are not tensors in general, there exists a coordinate system in which all components of the connection coefficients become zero and such
a coordinate system is called an \textit{affine coordinate}.
From the definition \eqref{a-connection}, we see that when $\alpha = 1$, all components of
$\Christo^{(1)}{}_{ijk}(\theta)$ vanish and when $\alpha = -1$,  those of $\Christo^{\star (-1) \; ijk} (\eta)$ vanish.
Consequently, the $\theta$ -coordinates are affine for the connection $\nabla^{(1)}$, and the $\eta$-coordinates are affine for $\nabla^{\star (-1)} $.
In this way, IG has the dually-flat structure \cite{AN,Amari}.

\subsection{Gradient-Flow Equations}
\label{GFeq}
For the sake of simplicity we consider the gradient-flow equations w.r.t a potential function $\Psi(\theta)$ or $\Psi^{\star}(\eta)$  in IG.
For more details, refer to  \cite{FA95,BN16} or \cite{WS24}.
The gradient-flow equations w.r.t. a convex  function $\Psi(\theta)$ are 
given by
\begin{align} 
\frac{d \theta^i}{d t} = 
g^{ij} ( \theta) \,
\frac{\partial \Psi(\theta)}{\partial \theta^j},
\label{theta-gradEq}
\end{align}
in the $\theta$-coordinate system.
By using the properties \eqref{g}, the left-hand side (LHS) of \eqref{theta-gradEq} is rewritten by
\begin{align}
\frac{d \theta^i}{d t} &= 
\frac{\partial \theta^i}{\partial \eta_j} 
\frac{d \eta_i}{d t} = 
g^{ij} ( \theta) 
\frac{d \eta_i}{d t},
\end{align}
and the RHS is
\begin{align} 
g^{ij} ( \theta) 
\frac{\partial \Psi(\theta)}{\partial \theta^j} = g^{ij} ( \theta) \, \eta_j.
\end{align}
Consequently, the gradient-flow equations \eqref{theta-gradEq} in the $\theta$-coordinate system are equivalent
to the linear differential equations
\begin{align}
\frac{d \eta_i}{d t} = \eta_i,
\label{LinearGradEq}
\end{align}
in the $\eta$-coordinate system. This linearization is one of the merits due to the dually-flat structures \cite{AN, Amari} in IG.
The other set of gradient-flow equations are given by
\begin{align} 
\frac{d \eta_i}{d t} = 
- g_{ij} ( \eta) 
\frac{\partial \Psi^\star(\eta)}{\partial \eta_j},
\label{eta-gradEq}
\end{align}
in the $\eta$-coordinate system.
Similarly, they are equivalent to the linear differential equations
\begin{align}
\frac{d \theta^i}{d t} = 
- \theta^i.
\end{align}
in the $\theta$-coordinate system.
 
Taking the derivative of both sides of \eqref{theta-gradEq} w.r.t. the parameter $t$ and using the relation \eqref{g} and $(\partial g^{i \ell}(\theta) / \partial \theta^k) g_{\ell j}(\theta) = -g^{i \ell}(\theta) \,  \partial g_{\ell j}(\theta) / \partial \theta^k$,
we obtain the  pre-geodesic equations \cite{W23}
\begin{align}
  \frac{d^2 \theta^i}{d t^2} + \Christo^{(-1) \; i}{}_{j k}(\theta) \frac{d \theta^j}{d t} \frac{d \theta^k}{d t}
 = \frac{d \theta^i}{d t}, \quad \Leftrightarrow \quad  \frac{d \theta^j}{dt} \nabla^{(-1)}_j  \frac{d \theta^i}{dt} = \frac{d \theta^i}{dt}, 
 \label{theta-pre-geodesic}
\end{align}
in the $\theta$-coordinate system. Here
\begin{align}
  \Christo^{(-1) \; i}{}_{j k}(\theta)  = g^{i\ell} (\theta) \Christo^{(-1)}{}_{\ell j k}(\theta) = g^{i\ell} (\theta) \frac{\partial g_{\ell j}(\theta) }{\partial \theta^k}
  =g^{i\ell} (\theta) \, C_{k \ell j}(\theta),
\end{align}
are the coefficients of the mixture ($\alpha \! = \! -1$) connection $\nabla^{(-1)}$.

Next, we introduce a positive scalar function $ \eta^2(\theta)$ of $\theta$ as
\begin{align}
  \eta^2(\theta) := g^{ij}(\theta) \frac{\partial \Psi(\theta)}{\partial \theta^i}  \frac{\partial \Psi(\theta)}{\partial \theta^j},
  \label{eta2}
\end{align}
which characterizes the rate of $\Psi(\theta)$ during the gradient-flow described by the $\theta$-gradient flow equations \eqref{theta-gradEq}. This can be seen as follows.
\begin{align}
  \frac{d \Psi(\theta)}{dt} = \frac{\partial \Psi(\theta)}{\partial \theta^i} \frac{d \theta^i}{dt} = g^{ij}(\theta) \frac{\partial \Psi(\theta)}{\partial \theta^i}  \frac{\partial \Psi(\theta)}{\partial \theta^j} = \eta^2(\theta),
  \label{dPsi_dt}
\end{align}
where we used Eq. \eqref{theta-gradEq}.

A similar argument applies to the other set of the gradient-flow equations \eqref{eta-gradEq} in the $\eta$-coordinate system.
Introducing another positive scalar function
\begin{align}
  \theta^2(\eta) := g_{ij}(\eta) \frac{\partial \Psi^{\star}(\eta)}{\partial \eta_i}  \frac{\partial \Psi^{\star}(\eta)}{\partial \eta_j},
  \label{theta2}
\end{align}
the corresponding relation of \eqref{dPsi_dt} are
\begin{align}
 - \frac{d \Psi^{\star}(\eta)}{d t} =  -\frac{\partial \Psi^{\star}(\eta)}{\partial \eta_i} \frac{d \eta_i}{d t} = -\theta^i \frac{d \eta_i}{d t} 
  = g_{ij}(\eta) \frac{\partial \Psi^{\star}(\eta)}{\partial \eta_i}  \frac{\partial \Psi^{\star}(\eta)}{\partial \eta_j} =  \theta^2(\eta).
\end{align}
Since $-\Psi^{\star}(\eta)$ is the entropy function, the LHS is the entropic rate in the entropic gradient-flow. Then it is found that this entropy rate 
$ - d \Psi^{\star}(\eta) / d t$ is characterized by the scalar field $\theta^2(\eta)$.

\section{Weyl symmetric action for the gradient-flow}
Here, we first briefly explain  the motion of a particle with the position dependent mass $m(x)$ in a curved spacetime in Weyl geometry.
The associated action $S_a$ \eqref{WSa} has Weyl symmetry.
Then we shall derive the gradient-flow equations in IG from the Euler-Lagrange (EL) equations of the Weyl symmetric action $S_a^{\rm IG}$ \eqref{SaIG},
which is analogous to  $S_a$ \eqref{WSa}.
The scalar function $\sqrt{\eta^2(\theta)}$ plays a similar role of the position dependent mass $m(x)$.

\subsection{Particle motion in curved spacetime and position dependent mass}
As mentioned in Introduction, a mass $m(x)$ is position dependent in the action $S_a$ which is invariant under the Weyl transformation \eqref{Weyl-transform}.  
Since we here consider a time-like case ($ ds^2 < 0$), we have
\begin{align*}
  \sqrt{-ds^2} = \sqrt{ -g_{\mu \nu} dx^{\mu} dx^{\nu}} =  \sqrt{ -g_{\mu \nu} \frac{dx^{\mu}}{d \lambda} \frac{ dx^{\nu}}{d \lambda } }d \lambda,
\end{align*}
for an arbitrary parameter $\lambda$, the invariant action $S_a$ can be written as follows.
\begin{align}
  S_a = -  \int  m(x)  \sqrt{ -ds^2 }
  = \int_{\lambda_i}^{\lambda_f} d \lambda \, L \left( x, \frac{d x}{d \lambda} \right).
  \label{WSa}
\end{align}
Here $L \left( x, \frac{d x}{d \lambda} \right)$ is the associated Lagrangian
\begin{align}
L \left( x, \frac{d x}{d \lambda} \right) = -  m(x)  \sqrt{ -g_{\mu \nu}(x) \frac{dx^{\mu}}{d \lambda} \frac{ dx^{\nu}}{d \lambda } }.
  \label{L}
\end{align}
It is convenient to rewrite this Lagrangian in a different way by introducing the so-called \textit{einbein field} $e(\lambda)$ \cite{einbein} as
\begin{align}
  L \left( x, \frac{d x}{d \lambda} \right) = \frac{ 1}{ 2 e(\lambda)} g_{\mu \nu}(x) \frac{dx^{\mu}}{d \lambda} \frac{ dx^{\nu}}{d \lambda }- \frac{ e(\lambda)}{2}  m^2(x).
  \label{einbein-form}
 \end{align}
The Euler-Lagrange (EL) equations w.r.t. $e(\lambda)$ are
\begin{align}
   0 &= \frac{\partial L}{\partial e}, \quad \Leftrightarrow \quad  e(\lambda) = \frac{1 }{ m(x)} \sqrt{-g_{\mu \nu}(x) \frac{dx^{\mu}}{d \lambda} \frac{ dx^{\nu}}{d \lambda}} 
   = \frac{1}{m(x)} \frac{\sqrt{-d s^2}}{d \lambda}.
   \label{e}
\end{align}
By substituting the expression of $e(\lambda)$ into \eqref{einbein-form}, we recover the expression \eqref{L} of the Lagrangian.
From \eqref{e}, we see that the quantity $e(\lambda) d \lambda$ is invariant against the change in the parameter $\lambda$.
In other words, the einbin transforms as $e(\lambda') = e(\lambda) \, (d \lambda / d\lambda')$ when the parameter changes
from $\lambda$ to $\lambda'$.

The canonical momentum $p_{\mu} := \partial L / \partial ( d x^{\mu} / d \lambda)$ and the EL equations are
\begin{subequations}
\label{ELeq}
\begin{empheq}[left=\empheqlbrace]{align}
    p_{\mu} &= \frac{1}{e(\lambda)} g_{\mu \nu}(x) \frac{d x^{\nu}}{d \lambda}, 
    \label{p} \\
   \frac{d p_{\mu}}{d \lambda}  &= \frac{1}{2 e(\lambda)} \frac{\partial g_{\nu \rho}(x)}{\partial x^{\mu}} \frac{d x^{\nu}}{d \lambda} \frac{d x^{\rho}}{d \lambda}
   - \frac{e(\lambda)}{2}  \frac{\partial m^2(x)}{\partial x^{\mu}}.
\end{empheq}
\end{subequations}
Eliminating $p_{\mu}$ leads to the geodesic equations:
\begin{align}
     \frac{d^2 x^{\mu}}{d \lambda^2} + \Gamma^{\mu}{}_{\nu \rho}(x)  \frac{d x^{\nu}}{d \lambda} \frac{d x^{\rho}}{d \lambda}
   = \frac{d  \ln e(\lambda)}{d \lambda} \frac{d x^{\mu}}{d \lambda}
   - \frac{e^2(\lambda)}{2}  g^{\mu \nu}(x) \frac{\partial m^2(x)}{\partial x^{\nu}}.
   \label{Geq}
\end{align}
From \eqref{e} and \eqref{p}, we have
\begin{align}
 p^2 = g^{\mu \nu}(x) p_{\mu} p_{\nu} =  \frac{g^{\mu \nu}(x)}{e^2(\lambda)} \frac{d x^{\mu}}{d \lambda} \frac{d x^{\nu}}{d \lambda} = - m^2(x).
 \label{on-shell}
\end{align}

\subsection{Derivation of the gradient-flow equations}

In conventional description, as explained in \ref{GFeq}, the gradient-flow in IG are formulated in the $N$-dimensional ($\theta$- or $\eta$-) space
with the time-evolutional parameter $t$. In order to relate the Weyl symmetry and the gradient-flow in IG, we extend this conventional formulation to an $N+1$ spacetime representation.

First we make correspondence between the variables in analytical mechanics and  those in IG as follows.
\begin{subequations}
\label{correspondence}
\begin{empheq}[left=\empheqlbrace]{align}
 \textrm{position: } \quad x^i  \quad &\Leftrightarrow \quad  \theta^i. \\
 \textrm{momentum: } \quad  p_i  \quad &\Leftrightarrow \quad  \eta_i,  \\
 \textrm{mass: } \quad m(x)  \quad &\Leftrightarrow \quad  \sqrt{\eta^2(\theta)}. \label{m2eta2}
\end{empheq}
\end{subequations}
We remind that the positive scalar function $\eta^2(\theta)$ is defined in \eqref{eta2}.
Introducing the line element $d \ell$ in the $N$-dimensional $\theta$-space by
\begin{align}
      d \ell^2 := g_{ij}(\theta) d\theta^i d\theta^j.
      \label{dell2}
\end{align}
We extend this $d \ell^2$ in $N$-dimensional space with the time evolutional parameter $t$ to $d s^2$ in $N+1$-dimensional spacetime as follows.
Since we consider a time-like case ($ ds^2<0$),  we assume, for a simplicity, that a static spacetime metric in the form
\begin{align}
      ds^2 = g_{\mu \nu}(\theta) d\theta^{\mu} d\theta^{\nu} = g_{0 0}(\theta) (d\theta^0)^2 + g_{ij}(\theta) d\theta^i d\theta^j = g_{0 0}(\theta) d\tau^2 + d \ell^2,
\end{align}
where we used \eqref{dell2}.
Here we choose $d\theta^0$  as the proper time $d\tau$, which is related with $ds^2 < 0$ as
\begin{align}
 d \theta^0 = d \tau  = \sqrt{-\eta^2(\theta) \, ds^2} > 0.
 \label{dtau}
 \end{align}
With the correspondence in \eqref{correspondence}, the on-shell relation \eqref{on-shell} leads to
\begin{align}
  g^{\mu \nu}(\theta) \eta_{\mu} \eta_{\nu} =  g^{00}(\theta) \eta_0^2 + g^{i j}(\theta) \eta_i \eta_j 
  = - \eta^2(\theta),
  \label{eta-square}
\end{align}
where $ g^{00}(\theta) = 1 /  g_{00}(\theta)$. 
Note that $\eta^2(\theta)$ should not be confused with $\eta^2 := g^{\mu \nu} \eta_{\mu} \eta_{\nu}$.
Then the associated Lagrangian becomes
\begin{align}
  L \left( \theta, \frac{d \theta}{d \lambda} \right) = \frac{ 1}{ 2 e(\lambda)} g_{\mu \nu}(\theta) \frac{d\theta^{\mu}}{d \lambda} \frac{ d\theta^{\nu}}{d \lambda }- \frac{ e(\lambda)}{2}  \eta^2(\theta),
  \label{LIG}
 \end{align}
 where the corresponding expression of the einbein $e(\lambda)$ is
 \begin{align}
   e(\lambda) = \frac{1 }{ \sqrt{\eta^2(\theta)}} \sqrt{-g_{\mu \nu}(\theta) \frac{d\theta^{\mu}}{d \lambda} \frac{ d\theta^{\nu}}{d \lambda}} 
   = \frac{1}{\sqrt{\eta^2(\theta)}} \frac{\sqrt{-d s^2}}{d \lambda}.
   \label{eIG}
\end{align}
 From \eqref{eta-square} and the definition \eqref{eta2}, we find
\begin{align}
  g^{00}(\theta) \eta_0^2   = - 2 \eta^2(\theta).
  \label{rel}
\end{align}
Let us choose the parameter $t$ such that the einbein $e(t)$ in \eqref{eIG} is unity, i.e., 
\begin{align}
   e(t) = \frac{1}{\sqrt{\eta^2(\theta)}} \frac{\sqrt{-d s^2}}{d t} = 1,
\end{align}
which means that $\sqrt{-ds^2} = \sqrt{\eta^2(\theta)} \, dt$.  Combining this relation with \eqref{dtau}, we have $d \tau = \eta^2(\theta) \, dt$.
Then the canonical momenta and EL equations of the Lagrangian \eqref{LIG} become
\begin{subequations}
\label{ELeqIG}
\begin{empheq}[left=\empheqlbrace]{align}
  \eta_0 &= g_{00}(\theta) \frac{d \tau}{d t} =g_{00}(\theta)  \, \eta^2(\theta), \quad  \eta_i = g_{ij}(\theta) \frac{d \theta^j}{d t},
  \label{EM1} \\
  \frac{d \eta_0}{d t} & = \frac{1}{2} \frac{\partial g_{\nu \rho}(\theta)}{\partial \tau} \frac{d \theta^{\nu}}{d t} \frac{d \theta^{\rho}}{d t}
   - \frac{1}{2}  \frac{\partial \eta^2(\theta)}{\partial \tau} = 0, \nonumber \\
    \frac{d \eta_i}{d t}  &= \frac{1}{2} \frac{\partial g_{\nu \rho}(\theta)}{\partial \theta^i} \frac{d \theta^{\nu}}{d t} \frac{d \theta^{\rho}}{d t}
   - \frac{1}{2}  \frac{\partial \eta^2(\theta)}{\partial \theta^i},
   \label{EM2}
\end{empheq}
\end{subequations}
where in the first equation in \eqref{EM2}, we used the fact that $g_{\nu \rho}(\theta)$ and $\eta^2(\theta)$ are independent of $\tau$.
It is remarkable that the second equations in \eqref{EM1} are equivalent to the gradient-flow equation \eqref{theta-gradEq}.

Next let us focus on the second equations in \eqref{EM2}.
From \eqref{rel} and the first equation in \eqref{EM1}, we find that
\begin{align}
  g_{00}(\theta) = - \frac{2}{ \eta^2(\theta) }, \quad \eta_0 = -2.
\end{align}
Then the first term in the RHS of the second equations in \eqref{EM2} are transformed as
\begin{align}
\frac{1}{2} \frac{\partial g_{\nu \rho}(\theta)}{\partial \theta^i} \frac{d \theta^{\nu}}{d t} \frac{d \theta^{\rho}}{d t} &= 
\frac{1}{2} \frac{\partial }{\partial \theta^i} \left(- \frac{2}{\eta^2(\theta)} \right) \left(\frac{d \tau}{d t} \right)^2 + \frac{1}{2} \frac{\partial g_{j k}(\theta)}{\partial \theta^i} \frac{d \theta^j}{d t} \frac{d \theta^k}{d t} \nonumber \\
&= \frac{\partial  \eta^2(\theta)}{\partial \theta^i}  + \frac{1}{2} \frac{\partial g_{j k}(\theta)}{\partial \theta^i} \frac{d \theta^j}{d t} \frac{d \theta^k}{d t}.
\end{align}
Consequently, the second equations in \eqref{EM2} are 
\begin{align}
\frac{d \eta_i}{d t}  = \frac{1}{2} \frac{\partial g_{j k}(\theta)}{\partial \theta^i} \frac{d \theta^j}{d t} \frac{d \theta^k}{d t}
   + \frac{1}{2}  \frac{\partial \eta^2(\theta)}{\partial \theta^i}.
   \label{eta2dt}
\end{align}
By differentiating the both sides of $\eta^2(\theta)$ in \eqref{eta2}, we have 
\begin{align}
  \frac{1}{2}  \frac{ \partial \eta^2 ( \theta) }{ \partial \theta^i} 
  &= \frac{1}{2} \frac{\partial g^{j k}(\theta)}{\partial \theta^i} \underbrace{\frac{\partial \Psi(\theta)}{\partial \theta^j} }_{\eta_j} \underbrace{ \frac{\partial \Psi(\theta)}{\partial \theta^k} }_{\eta_k} + g^{j k}(\theta)  \underbrace{ \frac{\partial^2 \Psi(\theta)}{\partial \theta^i \partial \theta^j}}_{g_{ij}(\theta)} \underbrace{\frac{\partial \Psi(\theta)}{\partial \theta^k}}_{\eta_k} .
  \end{align}
  By using the second equation in \eqref{EM1}, it follows
  \begin{align}
  \frac{1}{2}  \frac{ \partial \eta^2 ( \theta) }{ \partial \theta^i} 
  &= \frac{1}{2} \underbrace{ \frac{\partial g^{j k}(\theta)}{\partial \theta^i}  g_{k m}(\theta) }_{-\partial_i g_{k m}(\theta) g^{j k}(\theta)} \frac{ d \theta^m}{d t}  g_{j \ell}(\theta) \frac{d \theta^{\ell}}{dt} +
  \underbrace{ g^{j k}(\theta)  g_{ij}(\theta) }_{\delta^k{}_i}  \eta_k \nonumber \\
  &= -\frac{1}{2} \frac{\partial g_{k m}(\theta)}{\partial \theta^i}  \underbrace{g^{j k}(\theta) g_{j \ell}(\theta)}_{\delta^k{}_{\ell}}  \frac{ d \theta^m}{d t}  \frac{d \theta^{\ell}}{dt} + \eta_i
   = -\frac{1}{2} \frac{\partial g_{k m}(\theta)}{\partial \theta^i}   \frac{ d \theta^m}{d t}  \frac{d \theta^{k}}{dt} + \eta_i \nonumber \\
   &= -\frac{1}{2} \frac{\partial g_{j k}(\theta)}{\partial \theta^i}   \frac{ d \theta^j}{d t}  \frac{d \theta^{k}}{dt} + \eta_i.
  \end{align} 
 Substituting this relation into \eqref{eta2dt} leads to
\begin{align}
  \frac{d \eta_i}{d t} = \eta_i,
\end{align}
which are the linearized equations \eqref{LinearGradEq}.
In this way, we can derive both gradient-flow equations \eqref{theta-gradEq} and \eqref{LinearGradEq} from the Lagrangian \eqref{LIG}.
The corresponding action
\begin{align}
  S_a^{\rm{IG}} = \int \sqrt{\eta^2(\theta)} \, \sqrt{-ds^2},
  \label{SaIG}
\end{align}
is invariant under the local rescaling of 
\begin{align}
    ds^2 \to \Omega^2(\theta) ds^2, \quad \textrm{and} \quad \eta^2(\theta) \to \eta^2(\theta)  / \Omega^2(\theta),
    \label{localRS}
\end{align}
with a positive scalar function $\Omega(\theta)$.
The previous work \cite{W23} showed that, for the gradient-flow equations \eqref{theta-gradEq} in the $\theta$-space, the Weyl gauge (or covector) field is found to be
\begin{align}
 \omega_{\mu} = -\partial_{\mu} \ln \eta^2(\theta) = \partial_{\mu} \phi(\theta),
 \label{omega}
 \end{align}
 which implies $\phi(\theta) = - \ln \eta^2(\theta)$.
As explained in \ref{app2}, this $\omega_{\mu}$ of \eqref{omega} is consistent with the equivalence of the autoparallel \eqref{WgeodesicEq} and geodesic \eqref{Geq} equations.  In this case of $\phi(\theta) = - \ln \eta^2(\theta)$ in Weyl integrable geometry, the local rescaling \eqref{localRS} are equivalent with the Weyl gauge transformation \eqref{Weyl-transform}, in
which we see $\varphi(\theta) = \sqrt{\eta^2(\theta)}$. Consequently the action \eqref{SaIG} has Weyl symmetry.

Next, from \eqref{ELeqIG}, we obtain the geodesic equations w.r.t. the parameter $t$ as
\begin{align}
     \frac{d^2 \theta^{\mu}}{d t^2} + \Gamma^{\mu}{}_{\nu \rho}(\theta)  \frac{d \theta^{\nu}}{d t} \frac{d \theta^{\rho}}{d t}
   = - \frac{1}{2}  g^{\mu \nu}(\theta) \frac{\partial \ln \eta^2(\theta)}{\partial \theta^{\nu}},
   \label{geodesics}
\end{align}
which are also the autoparallel equations in the Weyl gauge of $(g_{\mu \nu}(\theta), \omega_{\mu} = -\partial_{\mu} \ln \eta^2(\theta) )$ as
  \begin{align}
  \frac{ d \theta^{\nu}}{d t} \overset{\textrm{\tiny{W}}}{\nabla}_{\nu}  \frac{ d \theta^{\mu}}{d t} = \frac{\partial_{\nu} \ln \eta^2(\theta)}{\partial \theta^{\nu}} \frac{d \theta^{\nu}}{d t}  \, \frac{d \theta^{\mu}}{dt}.
  \label{autoP}
 \end{align}


 
Now let us confirm the parameter $\tau$ is the proper time in the Weyl gauge $(g_{\mu \nu}, \omega_{\mu} = -\partial_{\mu} \ln \eta^2(\theta))$ in the following. The square $u^2(\tau)$ of the velocity vector $u^{\mu}(\tau)$ is
\begin{align}
  u^2(\tau) := g_{\mu \nu}(\theta) \frac{d \theta^{\mu}}{d \tau} \frac{d \theta^{\nu}}{d \tau}  
  =  \underbrace{g_{00} (\theta)  \left( \frac{d \tau}{d \tau} \right)^2}_{-2 / \eta^2(\theta)} + \underbrace{g_{ij}(\theta) \frac{d \theta^i}{d \tau}  \frac{d \theta^j}{d \tau} }_{1/\eta^2(\theta)} = - \frac{1}{\eta^2(\theta)},
\end{align}
then
\begin{align}
  \frac{ du^2(\tau)}{d \tau} =  \frac{1}{\eta^4(\theta)} \frac{d \eta^2(\theta)}{d \tau}  
  = - \frac{1}{\eta^2(\theta)} \left( - \partial_{\mu} \ln \eta^2(\theta) \right)  \frac{d \theta^{\mu}}{d \tau}  
= u^2(\tau) \, \omega_{\mu}  \frac{d \theta^{\mu}}{d \tau}.
\end{align}
Hence, from the definition \eqref{Proptime}, we see that the parameter $\tau$ is the Weyl proper time in the Weyl gauge $(g_{\mu \nu}, \omega_{\mu} = -\partial_{\mu} \ln \eta^2(\theta))$.

For the parameter $\tau$, the einbein is
$e(\tau) = 1 / \eta^2(\theta)$.
Then the corresponding geodesic equations \eqref{geodesics} become
\begin{align}
     \frac{d^2 \theta^{\mu}}{d \tau^2} + \Gamma^{\mu}{}_{\nu \rho}(\theta)  \frac{d \theta^{\nu}}{d \tau} \frac{d \theta^{\rho}}{d \tau}
   + \frac{d  \ln \eta^2(\theta)}{d \tau} \frac{d \theta^{\mu}}{d \tau}
   + \frac{1}{2 \eta^2(\theta)}  g^{\mu \nu}(\theta) \frac{\partial \ln \eta^2(\theta)}{\partial \theta^{\nu} } = 0,
   \label{EqA}
\end{align}
which are written in the form
\begin{align}
  \frac{d \theta^{\nu}}{d \tau} \Wnabla_{\nu} \frac{d \theta^{\mu}}{d \tau} = 0.
  \label{EqB}
\end{align}
Here the Weyl connection $\Wnabla$ is the Levi-Civita connection $ \overset{ \tilde{g} }{\nabla}$ w.r.t. the scaled metric $\tilde{g}_{\mu \nu}(\theta) = \eta^2(\theta) \, g_{\mu \nu} (\theta)$
since  when $\omega_{\rho} = -\partial_{\rho} \ln \eta^2(\theta)$, the scalar field $\phi(\theta) = -  \ln \eta^2(\theta)$ and  the coefficients \eqref{Wconnection} are
\begin{align}
 \WChrist{}^{\mu}{}_{\nu \rho}(\theta)  &=   \Christo^{\mu}{}_{\nu \rho} (\theta)
 +  \frac{1}{2} \Big( \delta^{\mu}_{\nu} \partial_{\rho} \ln \eta^2(\theta) + \delta^{\mu}_{\rho} \partial_{\nu} \ln \eta^2(\theta) -g^{\mu \sigma} \partial_{\sigma} \ln \eta^2(\theta)  g_{\nu \rho} \Big) \nonumber \\
 & = \frac{ g^{\mu \sigma}(\theta)}{2} \Big( \partial_{\rho} \,  g_{\nu \sigma}(\theta) + \partial_{\nu} \, g_{\sigma \rho}(\theta)  - \partial_{\sigma} \,  \tilde{g}_{\nu \rho}(\theta)  \Big) \nonumber \\
  & \qquad +  \frac{g^{\mu \sigma}(\theta)}{2 \eta^2(\theta)} \Big( g_{\nu \sigma} (\theta)\partial_{\rho} \eta^2(\theta) + g_{\sigma \rho}(\theta) \partial_{\nu} \eta^2(\theta) - g_{\nu \rho} (\theta)\, \partial_{\sigma} \eta^2(\theta)   \Big) \nonumber \\
 &= \frac{\tilde{g}^{\mu \sigma} (\theta)}{2} \Big( \partial_{\rho} \,  \tilde{g}_{\nu \sigma}(\theta) + \partial_{\nu} \, \tilde{g}_{\sigma \rho} (\theta)  - \partial_{\sigma} \,  \tilde{g}_{\nu \rho} (\theta) \Big)
 = \overset{ \tilde{g} }{\Christo}{}^{\mu}{}_{\nu \rho} (\theta).
 \end{align}
 It is worth note that Eq. \eqref{EqB} are also the geodesic equations $(d \theta^{\nu} / d \tau) \overset{\tilde{g}}{\nabla}_{\nu} (d \theta^{\mu} / d \tau) = 0$  in the Einstein gauge $(\tilde{g}_{\mu \nu}, \omega_{\mu} = 0)$,  in which
 the parameter $\tau$ is of course the proper time.

\section{Conclusions}
We have studied the gradient-flow equations in IG from the perspective of a time-like particle in a curved spacetime.
We have derived the gradient-flow equations in IG, from the action $S_a^{\rm{IG}}$  \eqref{SaIG} which is invariant under the Weyl transformation \eqref{Weyl-transform}. As explained in \ref{app1}, in Weyl integrable geometry the geodesic equations of the action $S_a^{\rm{IG}}$ coincide with the autoparalell equations in the Weyl gauge $(g_{\mu \nu}(\theta), \omega_{\mu} = -\partial_{\mu} \ln \eta^2(\theta))$. Recall that the Weyl symmetry is defined by the equivalent classes of $(g_{\mu \nu}, \omega_{\mu})$, and
the different choices of $(g_{\mu \nu}, \omega_{\mu})$ lead to the different Weyl gauges.
We have shown, from the invariant action $S_a^{\rm IG}$ \eqref{SaIG}, that
 in the Weyl gauge of $(g_{\mu \nu}(\theta), \omega_{\rho} = -\partial_{\rho} \ln \eta^2(\theta))$ and the parameter
 $t$, the gradient-flow eqs. \eqref{theta-gradEq} in $\theta$-space are obtained.
The parameter $t$ is not affine but related to the Weyl proper time $\tau$ as $d \tau = \eta^2(\theta) \, dt$.
The corresponding Weyl invariant connection $\Wnabla$ is equal to the Levi-Civita connection $ \overset{ \tilde{g} }{\nabla}$
w.r.t. the scaled metric $\tilde{g}_{\mu \nu}(\theta) = \eta^2(\theta) \, g_{\mu \nu}(\theta)$.

Since the autoparallel equations \eqref{autoP} in Weyl integrable geometry lead to the gradient-flow equations \eqref{theta-gradEq},
the space ($i$-th) components of \eqref{autoP} describe the same time-like curve $\theta^i = \theta^i(t)$ described by the pre-geodesic equations of the gradient-flow in IG, i.e.,
  \begin{align}
  \frac{ d \theta^j}{d t} \overset{\textrm{\tiny{W}}}{\nabla}_j  \frac{ d \theta^i}{d t} = \frac{\partial \ln \eta^2(\theta)}{\partial \theta^j}  \, \frac{d \theta^j}{d t} \frac{d \theta^i}{dt}, \quad
  \Leftrightarrow \quad
 \frac{d \theta^j}{dt} \nabla^{(-1)}_j  \frac{d \theta^i}{dt} = \frac{d \theta^i}{dt}.
 \end{align}
As a result, the Weyl connection $\Wnabla$ is related with the $\alpha$-connection of $\alpha=-1$ as
 \begin{align}
   \WChrist{}^i{}_{jk} \; \frac{d \theta^j}{d t} \frac{d \theta^k}{d t} 
 =
  {\Christo^{(-1)}}^i{}_{jk}(\theta) \; \frac{d \theta^j}{d t} \frac{d \theta^k}{d t}
  +  \frac{\partial \ln \eta^2(\theta)}{\partial \theta^j} \frac{d \theta^j}{d t} \frac{d \theta^i}{d t} - \frac{ d \theta^i}{dt}.
  \label{Walpha}
  \end{align}

On the other hand,
in the Einstein gauge $( \tilde{g}_{\mu \nu}(\theta) = \eta^2(\theta) \, g_{\mu \nu}(\theta),  \tilde{\omega}=0)$,
the autoparallel equations in Weyl integrable geometry are equivalent to the geodesic equations for the Levi-Civita connection $ \overset{ \tilde{g} }{\nabla}$ as
 \begin{align}
  \frac{ d \theta^{\nu}}{d \tau} \overset{\textrm{\tiny{W}}}{\nabla}_{\nu}  \frac{ d \theta^{\mu}}{d \tau} = 0, \quad
  \Leftrightarrow \quad
 \frac{d \theta^{\nu}}{d \tau} \overset{\tiny{\tilde{g}}}{\nabla}_{\nu} \frac{d \theta^{\mu}}{d \tau} = 0.
 \end{align}
 In this way, we also relate the geodesic equations in the Riemannian geometry equipped with the scaled metric 
 $\tilde{g}_{\mu \nu}(\theta) = \eta^2(\theta) \, g_{\mu \nu}(\theta)$ to the gradient-flow equations \eqref{theta-gradEq} in IG in the $\theta$-space, in whose dual $\eta$-space, the gradient-flow equations are linearized as shown in \eqref{LinearGradEq}. 
 
 Note that a similar argument applies to the gradient-flow equations \eqref{eta-gradEq} in the $\eta$-coordinate system, because due to the duality
 between $\theta$- and $\eta$-coordinate systems in IG. For example, the dual ($\eta$-) relation of \eqref{Walpha} relates the $\eta$-version
 of the Weyl connection to the $\alpha$-connection $\nabla^{(1)}$ with $\alpha=1$.
  
We hope that our result open up some connections to the fields that have not been considered before in the fields of IG.
Since we extended the conventional ($N$-space) gradient-flow in IG to the geodesic flow in $N+1$ spacetime, we can expect it to be related
to gravity theories and/or black hole physics, in which Weyl symmetry plays a key role.

\section*{Acknowledgements}
We acknowledge Akio Hosoya and Masahiro Morikawa for their interest in our work and their valuable comments.
The first named author (T.W.) was supported by Japan Society for the Promotion of Science (JSPS) Grants-in-Aid for Scientific Research (KAKENHI) Grant No. 22K03431 and No. 25K07110. The second named author (S.N.) was supported by JSPS KAKENHI Grant No. 24K17053.

\appendix
\section{Derivation of the autoparallel equations}
\label{app1}

We here provide the derivation of the most general expression \eqref{WgeodesicEq} of the autoparallel equations.
We  follow the similar way to obtain Eq. (17) in \cite{PS14}.
The starting point is
\begin{align}
   u^{\nu}(\lambda) \nabla_{\nu} u^{\mu}(\lambda) = f u^{\mu}(\lambda),
   \label{APeq}
\end{align}
where $f$ denotes a certain function of $u^{\mu}(\lambda)$. We can solve for $f$ as follows.
Using the non-metricity \eqref{non-metricity} and  $u^{\mu}(\lambda) := d x^{\mu} / d \lambda$ , we have
\begin{align}
  u^{\mu}(\lambda) \nabla_{\nu} u_{\mu}(\lambda) &= u^{\mu}(\lambda) \nabla_{\nu} (g_{\mu \gamma}(x)  u^{\gamma}(\lambda) ) = u^{\mu}(\lambda) \left\{   u^{\gamma}  (\lambda) \nabla_{\nu} g_{\mu \gamma}(x)  +
  g_{\mu \gamma}(x)  \nabla_{\nu} u^{\gamma} \right \} \nonumber \\
  &= 
  u^{\mu}(\lambda)  u_{\mu}(\lambda) \omega_{\nu} +
  u_{\gamma}(\lambda)  \nabla_{\nu} u^{\gamma}(\lambda).
\end{align}
Rearranging this as
\begin{align}
  u^{\mu}(\lambda) \nabla_{\nu} u_{\mu}(\lambda) - u_{\gamma}(\lambda)  \nabla_{\nu} u^{\gamma} (\lambda)=
  u^2(\lambda) \, \omega_{\nu}.
  \label{1}
\end{align}
By differentiating $u^2(\lambda)$ w.r.t. $x^{\nu}$ we have
\begin{align}
  u^{\mu}(\lambda) \nabla_{\nu} u_{\mu}(\lambda) + u_{\gamma}(\lambda)  \nabla_{\nu} u^{\gamma}(\lambda) =
  \partial_{\nu} u^2(\lambda).
  \label{2}
\end{align}
Subtracting \eqref{1} from \eqref{2}  and dividing by $2$ leads to
\begin{align}
  u_{\mu}(\lambda) \nabla_{\nu} u^{\mu}(\lambda) =  \frac{1}{2} \partial_{\nu}  u^2(\lambda)  - \frac{1}{2}   u^2(\lambda) \omega_{\nu}.
\end{align}
Contracting this relation with $u^{\nu}(\lambda)$ and using \eqref{APeq}, we have
\begin{align}
  f u^2(\lambda) & = u^{\nu}(\lambda) u_{\mu}(\lambda) \nabla_{\nu} u^{\mu}(\lambda) =  \frac{1}{2} u^{\nu}(\lambda)(\lambda) \partial_{\nu} u^2(\lambda) - \frac{1}{2}   u^2(\lambda) u^{\nu}(\lambda) \omega_{\nu}
  \nonumber \\
 & =  \frac{1}{2} \left( \frac{d  u^2(\lambda) }{d \lambda} - u^2(\lambda) u^{\nu}(\lambda) \omega_{\nu} \right),
\end{align}
where in the last step we used
\begin{align}
 u^{ \nu } (\lambda) \partial_{\nu} = \frac{ d \theta^{\nu}}{d \lambda} \frac{\partial}{\partial \theta^{\nu}}  = \frac{d}{d \lambda}.
\end{align}
Hence we obtain
\begin{align}
 f =  \frac{1}{2 u^2(\lambda) } \left( \frac{d  u^2(\lambda) }{d \lambda} - u^2(\lambda) u^{\nu}(\lambda) \omega_{\nu} \right).
\end{align}
By substituting this expression into \eqref{APeq}, we obtain \eqref{WgeodesicEq}.

\section{The equivalence of the autoparallel and geodesic equations}
\label{app2}
It is well known that for Levi-Civita connections in Riemannian geometry, the geodesics, which describe "straightest" curve, coincides with the autoparallels, which describe  "shortest" curves, whereas
for a general connection, the autoparallels  don't coincide with the geodesics.
In Weyl geometry, however, the autoparallels coincide with the geodesics \cite{Quiros23}.

For notational simplicity, let us denote the parameter $ \sqrt{-ds^2}$ as  $\mathfrak{s}$. From \eqref{e}, the einbein $e(\mathfrak{s})$ is
$e(\mathfrak{s}) = 1 / m(x)$.
Then the corresponding geodesic equations \eqref{geodesics} become
\begin{align}
     \frac{d^2 x^{\mu}}{d \mathfrak{s}^2} + \Gamma^{\mu}{}_{\nu \rho}(x)  \frac{d x^{\nu}}{d \mathfrak{s}} \frac{d x^{\rho}}{d \mathfrak{s}}
   &= \underbrace{\frac{d  \ln e(\mathfrak{s})}{d \mathfrak{s}} }_{ - \partial_{\nu}  \ln m(x)   \frac{d x^{\nu}}{d \mathfrak{s}} } \frac{d x^{\mu}}{d \mathfrak{s} }
   - \underbrace{\frac{e^2(\mathfrak{s})}{2}  \frac{\partial m^2(x)}{\partial x^{\nu} } }_{- \partial_{\nu}  \ln m(x) } g^{\mu \nu}(x)   \nonumber \\
   &= - \partial_{\nu}  \ln m(x)  \left( \frac{d x^{\nu}}{d \mathfrak{s}} \frac{d x^{\mu}}{d \mathfrak{s}}
   + g^{\mu \nu}(x)  \right).
   \label{GeoEq}
\end{align}

Recall that in Weyl integrable geometry, a Weyl covector $\omega_{\rho}$ is a gradient of a scalar function. We set $\omega_{\rho} = \partial_{\rho} \phi(x)$ with a scalar function $\phi(x)$.
For the parameter $\mathfrak{s}$, we have
\begin{align}
  u^2( \mathfrak{s} ) = g_{\mu \nu} \frac{d x^{\mu}}{d \mathfrak{s}} \frac{d x^{\nu}}{d \mathfrak{s}} = \frac{ds^2}{ d \mathfrak{s}^2} = -1.
\end{align}
Then the coefficients of the Weyl connection become
\begin{align*}
 \WChrist{}^{\mu}{}_{\nu \rho} \frac{d x^{\nu}}{d \mathfrak{s}}  \frac{d x^{\rho}}{d \mathfrak{s}} &=   \Christo^{\mu}{}_{\nu \rho} \frac{d x^{\nu}}{d \mathfrak{s}}  \frac{d x^{\rho}}{d \mathfrak{s}} 
 - \partial_{\nu} \phi(x) \frac{d x^{\nu}}{d \mathfrak{s}}  \frac{d x^{\mu}}{d \mathfrak{s}}
+  \frac{1}{2}g^{\mu \sigma} \partial_{\sigma} \phi(x) \underbrace{ g_{\nu \rho} \frac{d x^{\nu}}{d \mathfrak{s}}  \frac{d x^{\rho}}{d \mathfrak{s}}}_{-1} \nonumber \\
& =  \Christo^{\mu}{}_{\nu \rho} \frac{d x^{\nu}}{d \mathfrak{s}}  \frac{d x^{\rho}}{d \mathfrak{s}} 
 - \partial_{\nu} \phi(x) \frac{d x^{\nu}}{d \mathfrak{s}}  \frac{d x^{\mu}}{d \mathfrak{s}}
-  \frac{1}{2}g^{\mu \sigma} \partial_{\sigma} \phi(x) .
 \end{align*}
The RHS in the autoparallel equations \eqref{WgeodesicEq} becomes
\begin{align}
-\frac{1}{2}  \partial_{\rho} \phi(x) \, \frac{d x^{\rho}}{d \mathfrak{s} }   \frac{d x^{\mu}}{d \mathfrak{s} }.
\end{align}
Then the autoparallel equations for the parameter $\mathfrak{s}$ are
\begin{align}
\frac{d^2 x^{\mu}}{d \mathfrak{s}^2} +  \Christo^{\mu}{}_{\nu \rho} \frac{d x^{\nu}}{d \mathfrak{s}}  \frac{d x^{\rho}}{d \mathfrak{s}} 
 = \frac{1}{2} \partial_{\nu} \phi(x) \left( \frac{d x^{\nu}}{d \mathfrak{s}}  \frac{d x^{\mu}}{d \mathfrak{s}} +  g^{\mu \sigma} \right).
 \label{APE}
 \end{align}
Hence the geodesic equations \eqref{GeoEq} and the autoparallel equations \eqref{APE} are equivalent if 
\begin{align}
 \frac{1}{2} \partial_{\nu} \phi(x) = - \partial_{\nu} \ln m(x), \quad \Leftrightarrow \quad 
 \omega_{\nu} = \partial_{\nu} \phi(x) = - \partial_{\nu} \ln m^2(x).
\end{align}

Note that this is consistent with the correspondence $m^2(x) \; \Leftrightarrow \; \eta^2(\theta)$ in \eqref{m2eta2}
and the Weyl gauge (or covetor) field $\omega_{\nu} = - \partial_{\nu} \ln \eta^2(\theta)$ \eqref{omega}, which is found in Ref. \cite{W23}.

\section*{Data availability}
No data was used for the research described in the article.

\end{document}